\documentclass[a4paper]{article}

\usepackage{INTERSPEECH2022}
\usepackage{xcolor}

\title{Using Rater and System Metadata to Explain Variance in the VoiceMOS Challenge 2022 Dataset}
\name{Michael Chinen$^1$, Jan Skoglund$^1$, Chandan K A Reddy$^1$, Alessandro Ragano$^2$, Andrew Hines$^2$}
\address{
  $^1$Google\\
  $^2$University College Dublin, School of Computer Science, Dublin, Ireland}
\email{\{mchinen,jks,chandanka\}@google.com, alessandro.ragano@ucdconnect.ie, andrew.hines@ucd.ie}

\begin{document}

\maketitle
\begin{abstract}
  Non-reference speech quality models are important for a growing number of applications. The VoiceMOS 2022 challenge provided a dataset of synthetic voice conversion and text-to-speech samples with subjective labels.  This study looks at the amount of variance that can be explained in subjective ratings of speech quality from metadata and the distribution imbalances of the dataset.  Speech quality models were constructed using wav2vec 2.0 with additional metadata features that included rater groups and system identifiers and obtained competitive metrics including a Spearman rank correlation coefficient (SRCC) of 0.934 and MSE of 0.088 at the system-level, and 0.877 and 0.198 at the utterance-level.  Using data and metadata that the test restricted or blinded further improved the metrics.  A metadata analysis showed that the system-level metrics do not represent the model's system-level prediction as a result of the wide variation in the number of utterances used for each system on the validation and test datasets.  We conclude that, in general, conditions should have enough utterances in the test set to bound the sample mean error, and be relatively balanced in utterance count between systems, otherwise the utterance-level metrics may be more reliable and interpretable.

\end{abstract}
\noindent\textbf{Index Terms}: speech quality estimation, reference-free, non-intrusive, MOS, metadata, raters

\section{Introduction}

Reference-free or non-intrusive speech quality models are generally more useful than intrusive models for applications where obtaining a reference is inconvenient, impractical, or non-existent.  The VoiceMOS challenge 2022~\cite{huang2022voicemos} used a dataset~\cite{cooper2021voices} that aggregated new raters on previous Blizzard Challenges and Voice Conversion speech data.  The challenges were concerned with text-to-speech (TTS) and voice conversion applications, which are historically inconvenient to obtain a reference for.  Non-intrusive models are desirable for a growing number of applications.  For example, while intrusive quality models are well-suited to traditional codecs such as Opus, recent neural codecs can be problematic for full-reference quality models.  Generative neural frameworks such as the Lyra codec~\cite{kleijn2018wavenet} and TacoTron~\cite{shen2018natural}, which use a WaveNet-based architecture~\cite{oord2016wavenet} produce output that is not typically aligned with the original reference signal, with natural-sounding deviations in pitch and time.

While deep learning has made huge advancements in most areas of speech synthesis and processing over the last decade, the progress of speech quality models due to deep learning has only recently been notable.  One of the reasons for this is the limited amount of subjective ratings data that is publicly available.  In the past few years several notable models have appeared, including DNSMOS~\cite{reddy2021dnsmos} and NISQA~\cite{mittag2019non}.  Both DNSMOS and NISQA are from Microsoft, which has managed to assemble a very large dataset on the order of 100k utterances, and has recently released a public subset of the data that may encourage more neural models.  Additionally, MOSNet~\cite{lo2019mosnet} is a deep learning-based MOS estimator for voice conversion.  Semi-supervised architectures such as one using clustering~\cite{ragano2021more} and SESQA~\cite{serra2021sesqa} provide useful methods for dealing with the quality label data shortage, and lightweight and interpretable models that work with smaller amounts of data~\cite{chinen2021speech}.  The advancements in transformer-based pretrained models such as wav2vec 2.0~\cite{baevski2020wav2vec} and HuBERT~\cite{hsu2021hubert} enable researchers to explore another semi-supervised method to take advantage of the large amount of speech data that exists without subjective labels.  SSL-MOS~\cite{cooper2021generalization} was one of the baselines provided by the challenge organizers which uses wav2vec 2.0 with a minimal extra layer with promising results.

In parallel to the deep learning improvements that are mostly driven by extracting more useful information from the waveform, researchers have made progress in obtaining a better understanding of the biases and factors of listening tests that are independent of the speech signal being rated~\cite{zielinski2008some}.  For example, research has found that a significant amount of bias may be attributable to properties of the listeners, including their language and culture, as well as their individual tendencies to rate high or low~\cite{mossavat_bayesian_modeling,chinen2021lanugage}.  LDNet~\cite{huang2021ldnet}, a baseline model for this challenge, considers rater metadata.

Alongside listener metadata, it may be valuable to consider using metadata related to the signal.  Previous research has used aspects of network transmission, such as packet loss rate~\cite{jelassi2012single, rungruangthum2015simple}.  Synthetic data has additional properties that can be used as metadata that do not exist in natural data, such as the synthesis system description.  If a model uses metadata such as system identifiers, is it only useful on systems seen during training, or can it make equivalent or improved predictions for unseen systems?  We explore this question and propose a method that allows using identifier metadata in such a way that it can make useful predictions about unseen systems as well.

In this paper, we build off of wav2vec 2.0 baseline systems to examine the effect of metadata, as well as the design choices for the challenge.  Our main contributions in this paper are:
\begin{itemize}
\item Determining the amount of error and correlation that can be explained by metadata predictors such as system and rater identifiers.
\item Analyzing systematic differences in the validation and test set that can explain the test results.
\item Finding that utterance-level metrics are more useful than system-level metrics due to the majority of the test systems having a very small number of utterances.
\end{itemize}

\section{Dataset and Feature Selection}

\begin{figure*}[t]
  \centering
  \includegraphics[width=\linewidth]{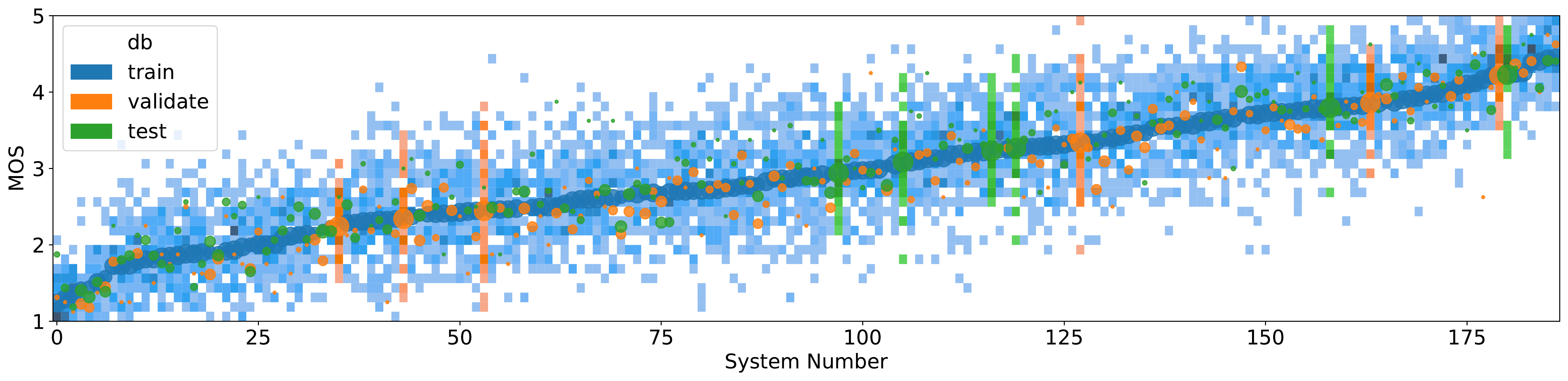}
  \caption{2D histogram of the MOS for utterances for each system that was unique to each dataset.  The validation dataset visualization excludes the systems that the training set also had in common, and the test set excludes the systems that either the validation or training set had in common.  The systems are ordered by MOS over all utterances.  The circles indicate the global MOS for each system on each dataset split, with size indicating the number of utterances.}
  \label{fig:system_histogram}
\end{figure*}

There are several interesting properties about the VoiceMOS Challenge 2022 dataset.  First the dataset is on synthetic data that does not have a reference.  Second, the synthesis system under test is identifiable from the file name.  Third, the rater's individual score and demographic information is provided.  Fourth, there were 304 total raters, but each rater always rated utterances with the same seven other raters (for a total of 38 rater groups), and each group rated each system exactly once.  
These properties will be discussed in detail with regards to modeling decisions.

Since the synthesis system can be identified from the file name, (which was allowed by the challenge for this purpose,) it is a potential input to the model.  Does it make sense to use the system metadata this way?  On the one hand, the mean scores among the systems (i.e. the system-wide aggregated MOS) are widely distributed as can be seen in Figure~\ref{fig:system_histogram}, so the MOS predictions from a model with the system identifier as its sole input should be expected to have some positive correlation with subjective MOS.  Since one of the challenge's metrics is system-aggregated error and correlation, adding a system predictor may reduce system error by pooling towards the system mean. On the other hand, in the extreme case where a model is able to perfectly predict the system from the signal, there would be no benefit to adding the system identifier as another feature once the signal is provided.  Additionally, adding the system information may introduce issues such as failure to generalize to out-of-sample data. The test set content is not significantly different from the training set for this challenge, so it is an interesting feature that we will consider.  If the system ID was used for real world applications and there is a question about the inference data not being represented by the training data, the 'blinded' (unknown) one-hot embedding could be used at inference.

In contrast to the system metadata, there is no causal link between the raters and the signal (i.e. it is not possible to predict anything about the rater given the signal alone).  Groups of raters are randomly assigned to signals.  For each rating, detailed information such as the rater's age, sex, and hearing ability is provided, along with a unique identifier that can be used to identify a certain rater over multiple ratings.  An individual rater always appeared in the same group of eight raters, which makes the individual rater details less important when considering the MOS averaged over all eight raters.  Some schemes could have taken advantage of this by augmenting the data by randomly withholding individual raters during training, but since the test ratings would also be averaged over the same eight raters there was not much of an incentive to do so, although it may help as regularization.  We selected a simple rater group one-hot encoded feature that identified the 36 different rater groups that were present in the train set.

\section{Model design}

The choice of input features and model architecture with respect to challenge objectives and dataset properties informed the model design.  The specific modeling decisions and rationale for them is discussed in this section.

\subsection{Input features}
Taking into consideration the dataset and challenge properties in the previous section, we constructed a model that used the following input features from the dataset:
\begin{itemize}
\item The speech waveform signal.
\item The rater group, encoded as a one hot vector.
\item The synthesis system, encoded as a one hot vector.
\item MOS predictions from the baseline SSL-MOS model with the small wav2vec 2.0.
\end{itemize}

The rater group and system features use the number of systems and rater groups seen during training, plus one 'unknown' class that is used for any unseen group.  During training, a process similar to dropout (without rescaling) is used to replace the known group with an 'unknown' class randomly at a certain probability.  This 'unknown' category can be used at inference for any unseen system or rater, as well as cases when the rater data is not available as in the test set.  This allows the model's subsequent layers to map the discrete identifiers to a continuous latent space for rater and systems that reflects their properties.  If a rater is usually optimistic in their ratings, the model should be able to learn this in training by predicting lower scores for average raters with the same utterance.

\subsection{Architectures}

wav2vec 2.0 was shown to obtain good baseline results for this challenge in SSL-MOS~\cite{cooper2021generalization}.  The speech waveform is input to wav2vec 2.0.  For wav2vec 2.0, the public (available on the 'fairseq' GitHub repository~\cite{ott2019fairseq}) 95M-parameter 'base' model pretrained on LibriSpeech was used, and an intermediate layer's 768-dimensional latent space with a variable number of frames was extracted.  Several convolutional and pooling layers remove the time dimension and reduce it down to 64-dimensions.  For both models with and without wav2vec 2.0, the one-hot metadata features described above are concatenated to the pooled wav2vec 2.0 64-dimensional features, and four additional fully connected layers reduce it to a 1-dimensional MOS.  This predicted MOS is used with an L1 loss on the subjective MOS.

\section{Experiment}

We conducted an experiment to test the effect of adding metadata to the model in a number of combinations.

Depending on the model, the training time was up to 3 hours on an Nvidia RTX 3090 GPU, with a learning rate of 0.001 on an Adam optimizer for 20-30 epochs.  A number of different combinations of the input features were selected as models to test.  The largest version of the system had 558k trainable parameters, with 95M additional frozen parameters from wav2vec 2.0.

\subsection{Design}

Since the rater information was not provided on the test set until after the challenge, results are provided using the 'blinded' rater group inputs (that is, using the 'unknown' one-hot value for all inferences,) or the actual rater group.  In both cases, the 'known' raters are still the ones in the train set only.

The experiments we ran were conducted after the challenge phase so that we could evaluate the results of many models on the test set and analyze their performance.  The instructions stated that the validation set should not be trained on when evaluating the test set.  Here we demonstrate what happens if we include the additional raters and systems from the validation set into the training set for the purposes of evaluating on the test set to see how much of an advantage that would bring.  To be clear, we do not have any reason to believe that any of the participants applied a cheating strategy, and there was no cash prize for this challenge; the purpose of this paper is to ascertain the amount of advantage that a model could obtain by training on the additional data and metadata.  Additionally, although our results are competitive with the challenge entries, they were chosen by taking the model checkpoint with the best performance on the test set, which was not possible to do during the challenge (only three submissions were allowed to be scored).  This allows the ablation studies to be more consistent, but these results are the product of peeking into the test set that the challenge rules would not allow for and should not be compared to challenge entries for this reason.

\subsection{Results}

The results for the challenge for a subset of the conditions under test are presented in Table~\ref{tab:results}. The SSL-MOS predictions provided the strongest baseline of all the features.  Additionally, the MSE for a constant mean predictor that predicts a MOS of 2.93 (the train set's aggregate MOS) for each utterance, is included as a no-input oracle baseline, as well as a no-input DNN that also produces the same MOS for all utterances to show the effect of batch size limitations compared to the oracle.

\begin{table}[th]
  \caption{Challenge results on the test set.  The model names indicate what features were used for training as follows. W2V: wav2vec 2.0, BR: blinded rater, R: known rater, S: synthesis system, M:baseline SSL-MOS w2v MOS. 'valtrain' indicates a 'cheating' approach where the model trained on both the train and validation sets. The 'Constant Mean' baseline model always predicts a MOS of 2.93 for any utterance.}
  \label{tab:results}
  \centering
  \begin{tabular}{ r | c c | c c }
    \toprule
          & \multicolumn{2}{c}{\textbf{System}} & \multicolumn{2}{c}{\textbf{Utterance}} \\
    Model & SRCC & MSE & SRCC & MSE \\

    \midrule
    Constant Mean               & -                & 0.671 & -     & 0.847 \\
    (No Input DNN)              & -                & 0.656 & -     & 0.852 \\
    R                           & 0.146            & 0.700 & 0.140 & 0.843 \\
    S                           & 0.882            & 0.189 & 0.792 & 0.357 \\
    S+BR                        & 0.878            & 0.190 & 0.787 & 0.347      \\
    S+R                         & 0.872            & 0.178 & 0.781 & 0.336     \\
    W2V                         & 0.914            & 0.165 & 0.841 & 0.273     \\
    W2V+BR                      & 0.908            & 0.212 & 0.849 & 0.299     \\
    W2V+R                       & 0.907            & 0.212 & 0.857 & 0.283     \\
    W2V+S                       & 0.906            & 0.201 & 0.849 & 0.284     \\
    W2V+BR+S                    & 0.906            & 0.109 & 0.855 & 0.224              \\
    W2V+R+S                     & 0.911            & 0.105 & 0.869 & 0.202 \\
    M                           & 0.921            & 0.148 & 0.878 & 0.196 \\
    BR+S+M                      & 0.933            & 0.085 & 0.877 & 0.202 \\
    R+S+M                       & 0.936            & 0.084 & 0.889 & 0.180 \\
    W2V+BR+S+M                  & \textbf{0.934}            & \textbf{0.088} & \textbf{0.877} & \textbf{0.198}   \\
    W2V+R+S+M                  & 0.935            & 0.085 & 0.887 & 0.179 \\
    W2V+BR+S+M (valtrain)      & 0.937            & 0.080 & 0.880 & 0.196   \\
    W2V+R+S+M (valtrain)       & 0.939            & 0.079 & 0.891 & 0.175   \\
    \bottomrule
  \end{tabular}
  
\end{table}

\section{Analysis and discussion}

These results are useful for analyzing the effect of including metadata in a challenge.  This section provides an analysis of the properties of the training, validation, and test splits of the dataset with respect to the result.  This will be useful for better understanding the results of this particular challenge.

\subsection{System analysis}

The train set had 175 unique systems for which the aggregated MOS was 2.93.  The validation set had the original systems but also had six additional unseen systems.  These six unseen systems had a MOS of 3.07.  The test set held an additional six systems which had a MOS of 3.42.  Because of the limited number of systems, the test set's distribution of unseen systems is quite different from the train set, as can be seen from Figure~\ref{fig:system_histogram}.  The validation and test datasets also contained the training systems, but with much fewer utterances than were present in the training set.

\begin{figure}[t]
  \centering
  \includegraphics[width=\linewidth]{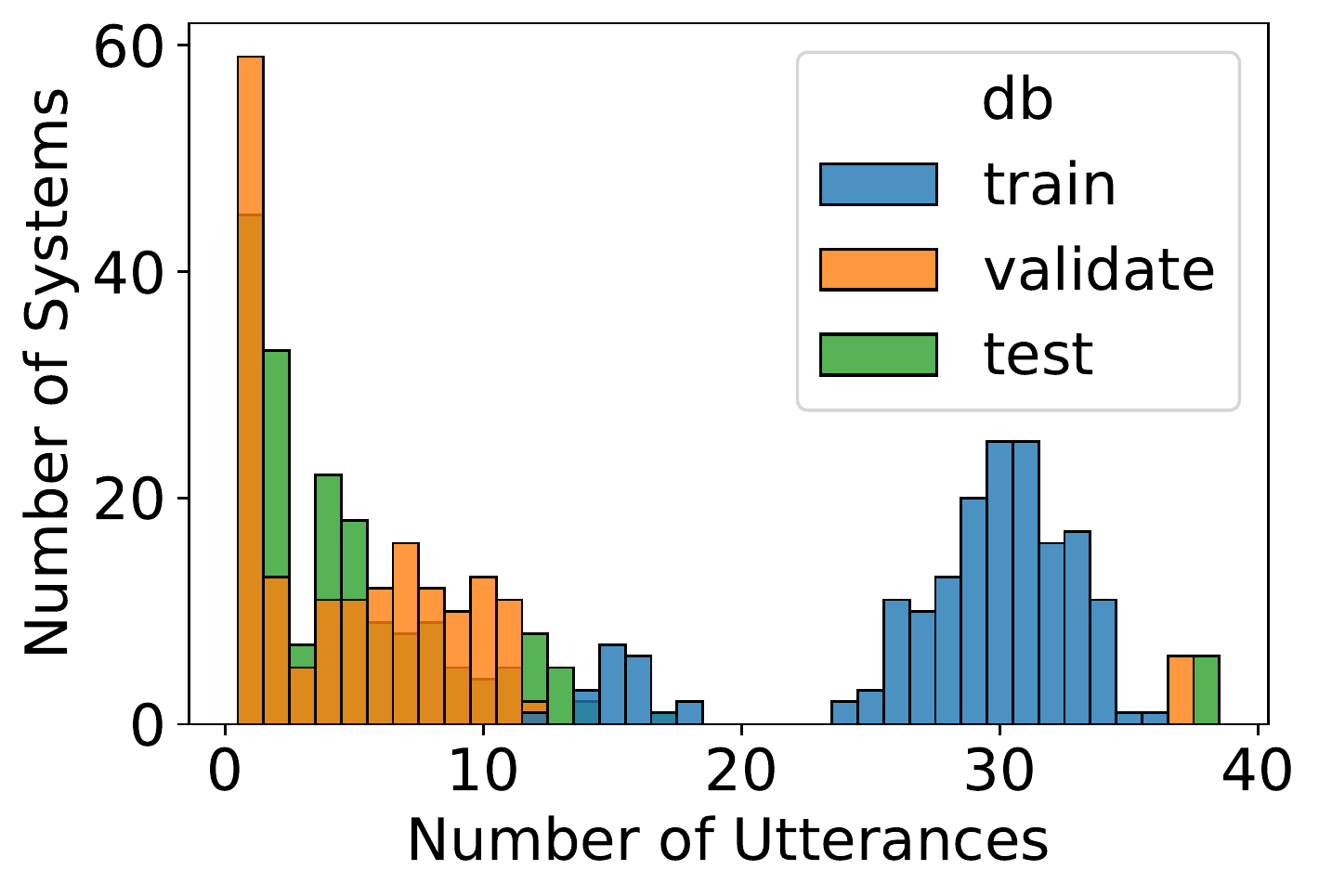}
  \caption{Histogram of number of utterances per system in each of the dataset splits.}
  \label{fig:utterance_count}
\end{figure}

We consider the meaning of system-level metrics for this dataset, which were the main metrics the challenge was graded on.  For all of our models, including metadata-only models without wav2vec 2.0, the system correlation for the training set was above 0.98, and as high as 0.997, which is significantly higher than the test set and might indicate that the model had problems such as overfitting.  However, it was noted that even when training on the validation set, the validation system SRCC was only around 0.96.  Further inspection revealed that a likely explanation for this is the small amount of utterances for each system in the validation and test sets (see Figure~\ref{fig:utterance_count}, resulting in many systems that have only one or two utterances, with a mean very far from the training system mean.  This is shown by the circles in Figure~\ref{fig:system_histogram}, which move closer to the train set's mean as they grow larger, illustrating the weak law of large numbers.  In particular, note that there are no large or medium circles that are the extreme outliers, while the small dots are exclusively the outliers, such as the one that represents system number 63 (system f7bda), with a test system mean of 3.875 with just one utterance, and a train MOS of 2.504, with 28 utterances.  If we consider the training set and test set to be two samples from the distribution of the system, the training set's system sample mean with 30 or more utterances should be tightly bound and is far more likely to be closer to the true mean than the sample mean of size one or two in the test set, which is expected to contain large errors.

Furthermore, since the system aggregation was not weighted by utterance count, these small systems had an outsized effect on the correlation and MSE that is misleading because it the system MOS for this dataset split will have a much higher error than other systems that have a large number of utterances.  More generally, unweighted condition-based MOS aggregation should be meaningful when the conditions have distinguishable MOS, with a reasonable number of utterances per condition to bound the error of the sample mean (see ITU-T P.800 on condition recommendations~\cite{itu1996p800}).  When this is not possible, utterance-level metrics such as SRCC and MSE would be more interpretable and useful as a measure of predictive power.

It is clear from the results that the synthesis system metadata contains a large amount of information about the quality. The model that only used system and rater information was able to have an utterance correlation of 0.787 on the test set, which is fairly large.  However, using system metadata on top of wav2vec 2.0 did not improve the metrics unless rater data was also included, so the majority of the system information is also present in the wav2vec 2.0 features.

\subsection{Rater analysis}

There were 36 distinct rater groups in the train set, which had a mean rating of 2.93.  The validation set had one additional unseen rater group with a MOS of 2.78.  The test set had one additional rater group with a MOS of 3.20.  Since there is only one additional rater unseen in the validation set, the possibility of the unseen raters not matching the distribution of the training set is relatively high.  In this particular dataset it would be slightly disadvantageous to train on the validation set for the purpose of learning more about the unseen raters because the test set's unseen rater happened to have a MOS above the train set's (as opposed to the validation set's unseen rater, which is below the mean).  However, this small disadvantage is likely overruled by the large advantage gained from training on new systems and utterances.  In the non-blinded case, only two rater groups (one from the test set and one from the validation set) used the 'unknown' dropout-like value that all rater groups were randomly trained with some percentage of the time.  In addition to the results, we found that removing the utterances with unknown raters from the unblinded best model increased the system SRCC from 0.939 to 0.957 on that subset of the test set, providing some evidence for this hypothesis.

The effect of adding a rater group feature during training appeared to have a positive effect, regardless of whether or not the test set raters were blinded.  wav2vec 2.0 MSE metrics were significantly improved when adding both system and rater, from 0.273 to 0.202 (utterance) and 0.165 to 0.105 (system).  Since this effect was only observed when both metadata were added, there may be an interaction effect (e.g. certain raters prefer certain systems) that requires both features.  

\subsection{Training on the validation set}
As the above system and rater analysis indicates, there are properties of the dataset split that produce advantages and disadvantages for by training on the validation dataset, which was prohibited by the rules of the challenge.  The results indicate that there is only a slight net advantage, as the per utterance metrics were slightly improved by several thousandths.  The validation and test sets had different distributions as observed above, and this likely limited its usefulness of training on the validation set with respect to inferring on the test set.
\section{Conclusion}

The VoiceMOS 2022 Challenge provided a rich dataset on voice conversion and text-to-speech synthesis data.  The dataset also had several properties that were interesting for exploring the effects of metadata and experimental and challenge design.  Our findings provide evidence that metadata can be a meaningful source of information even when the test set is blinded, because the models can explain some of the variance of the scores with this data that it would not be able to do with the signal alone.  More generally, besides synthesis system and rater information, metadata about the conditions under test (e.g. codec, bitrate, recording equipment) or more information about the test (e.g. date, location, rater screening process) could be improve the model using the dropout technique demonstrated to improve a general model that has option of using the metadata at inference only if it is available.  Lastly, the design of the splitting of the dataset can create interesting problems, such as the system-level aggregation issues discussed above.  This is a complicated problem because there are often limited amounts of data to begin with, which is usually the case for speech quality applications.  Metadata analysis can help to detect and explain these sorts of issues.  As the predictive power of no-reference models grows due to better signal analysis, organizers should continue offering more application-specific metadata, as was done for this challenge.

\section{Acknowledgements}

We would like to thank the VoiceMOS challenge organizers for their efforts in creating this challenge, and for providing the test set metadata after the challenge was finished.

\bibliographystyle{IEEEtran}

\bibliography{mybib}

\end{document}